\documentclass[pra,aps,showpacs,superscriptaddress,twocolumn]{revtex4}

\usepackage{amsmath}
\usepackage{amsfonts}
\usepackage{bm}
\usepackage{graphicx}

\begin{document}

\title{Energy shift of magnons in a ferromagnetic spinor-dipolar
Bose-Einstein condensate}

\author{Hiroki Saito}
\affiliation{Department of Engineering Science, University of
Electro-Communications, Tokyo 182-8585, Japan}

\author{Masaya Kunimi}
\affiliation{Department of Engineering Science, University of
Electro-Communications, Tokyo 182-8585, Japan}

\date{\today}

\begin{abstract}
Motivated by the recent experiment performed by the Berkeley group
[G. E. Marti {\it et al.}, Phys. Rev. Lett. {\bf 113}, 155302 (2014)],
we consider the dynamics of magnons in a spin-1 spinor-dipolar
Bose-Einstein condensate, using mean-field theory.
We show that the effective mass of a magnon is increased by the magnetic
dipole-dipole interaction, as observed in the experiment.
The magnon mass is also decreased by changing the direction of the
magnetic field.
The increase and decrease in the magnon mass manifest themselves in the
acceleration of the magnons.
\end{abstract}


\maketitle

Due to the long-range and anisotropic properties of the magnetic
dipole-dipole interaction (MDDI), Bose-Einstein condensates (BECs) of
ultracold atoms with a magnetic dipole moment exhibit a variety of
intriguing phenomena in experiments, including anisotropic deformation and
excitation of the condensate~\cite{Stuhler,Bismut,Pollack,Lu}, instability
in a prolate system~\cite{Lahaye07,Koch}, {\it d}-wave collapse and
expansion~\cite{Lahaye08,Metz,Aikawa}, and spinor-dipolar
dynamics~\cite{Pasquiou,Eto}.
Many theoretical investigations have also been made of
supersolidity~\cite{Goral}, the roton
spectrum~\cite{ODell,Santos,Wilson,Cherng}, rotating
properties~\cite{Cooper,Zhang,Bijnen,Klawunn,Shirley}, soliton
stability~\cite{Pedri,Tikho,Nath}, the Einstein-de Haas
effect~\cite{Kawaguchi_edh,Gawryluk,Sun}, spin-texture
formation~\cite{Yi06,Kawaguchi06,Takahashi,Hoshi,Huhtamaki}, Rosensweig
instability~\cite{Saito09}, and various spin
dynamics~\cite{Yasunaga,Huhtamaki11}.

Observation and identification of dipolar effects in a spin-1
$^{87}{\rm Rb}$ BEC are difficult~\cite{Vengalattore,Kawaguchi10} compared
with the cases of $^{52}{\rm Cr}$, $^{164}{\rm Dy}$, and $^{168}{\rm Er}$
BECs, since the MDDI of spin-1 $^{87}{\rm Rb}$ atoms is much smaller than
the characteristic energies in the system.
Recently, using the technique of precise {\it in situ} measurement of spin
states, the Berkeley group observed magnon dynamics in a spin-1
$^{87}{\rm Rb}$ BEC~\cite{Marti}.
Although the mean-field theory without MDDI predicts that the dispersion
relation of a magnon is the same as that of a free atom~\cite{Ho,Ohmi},
the experimental result significantly deviates from the prediction and the
effective mass of a magnon is measured to be $\simeq$ 3\% larger than that
of a free atom.
This deviation is much larger than the quantum many-body
correction~\cite{Phuc}.
The authors of Ref.~\cite{Marti} attribute the deviation to the MDDI,
which has yet to be evaluated by quantitative theoretical analysis.

In this Rapid Communication,
we investigate the effect of the MDDI on the
dynamics of a magnon in the same setup as was used by the Berkeley
experiment~\cite{Marti}.
Using the mean-field theory with the MDDI, we confirm that the dispersion
relation and the effective mass of a magnon deviate from those of a free
atom, as observed in the experiment.
We also propose a simple modification to the experiment:
when the direction of the magnetic field is changed from the long axis to
the tight axis of the oblate condensate, the magnon mass is found to be
decreased.
We show that the increase and decrease in the magnon mass affect the
acceleration of magnons in the magnetic field gradient.

We employ the mean-field approximation to study the dynamics of a spin-1
BEC with MDDI.
In the experiment~\cite{Marti}, a magnetic field of 115 mG is applied
(we take its direction as the $z$ axis), which induces the Larmor
precession at a frequency of $\simeq 80$ kHz.
This time scale is much faster than that of the condensate dynamics, and
therefore the MDDI can be time-averaged with respect to the Larmor
precession~\cite{Giovanazzi,Kawaguchi07}.
The macroscopic wave functions $\psi_m(\bm{r}, t)$ for the magnetic
sublevels $m = \pm 1$ and 0 thus obey the nonlocal Gross-Pitaevskii (GP)
equation given by
\begin{eqnarray} \label{GP}
i \hbar \frac{\partial \psi_m}{\partial t} & = & \left(
-\frac{\hbar^2}{2M} \nabla^2 + V + g_0 \rho + m^2 q \right) \psi_m 
\nonumber \\
& & + g_1 \bm{F} \cdot \sum_{m'} \bm{f}_{mm'} \psi_{m'}
\nonumber \\
& & + g_d \int d{\bm r}' \frac{1 - 3 \cos^2 \theta}{|\bm{r} - \bm{r}'|^3}
\nonumber \\
& & \times \sum_{m'} \left[ 3 F_z(\bm{r}') f^z_{mm'} - \bm{F}(\bm{r}') \cdot
\bm{f}_{mm'} \right] \psi_{m'}(\bm{r}),
\nonumber \\
\end{eqnarray}
where $M$ is the mass of an $^{87}{\rm Rb}$ atom, $V = M (\omega_x^2 x^2 +
\omega_y^2 y^2 + \omega_z^2 z^2) / 2$ is the harmonic trap potential,
$\rho = \sum_m |\psi_m|^2$ is the total density, $q$ is the quadratic
Zeeman shift, $\bm{f}$ is the vector of spin-1 matrices, $\bm{F} =
\sum_{mm'} \psi_m^* \bm{f} \psi_{m'}$ is the spin density, and $\theta$ is
the angle between $\bm{r} - \bm{r}'$ and the $z$ axis.
The interaction coefficients in Eq.~(\ref{GP}) are defined as $g_0 = 4 \pi
\hbar^2 (a_0 + 2 a_2) / (3M)$, $g_1 = 4 \pi \hbar^2 (a_2 - a_0) / (3M)$,
and $g_d = \mu_0 (\mu_B / 2)^2 / (8 \pi)$, where $a_S$ is the $s$-wave
scattering length for two colliding atoms with total spin $S$, $\mu_0$ is
the vacuum permeability, and $\mu_B$ is the Bohr magneton.
The trap frequencies are taken to be $(\omega_x, \omega_y, \omega_z) /
2\pi = (9, 300, 4)$ Hz, as in the experiment.
The wave function is normalized as $\int \rho d\bm{r} = N$, where $N$ is
the number of atoms.
We take $N = 3.2 \times 10^6$ so that the peak density at the center of
the trap becomes the experimental value, $1.5 \times 10^{14}$
${\rm cm}^{-3}$.
The quadratic Zeeman shift is $q \simeq h \times 1$ Hz for the magnetic
field of 115 mG, which does not affect the following results.

The initial state is the ground state of the $m = -1$ component,
$\psi_{-1} \equiv \Psi_{-1}$, with $\psi_1 = \psi_0 = 0$.
Magnons with wave numbers $\pm K$ are excited as
\begin{equation} \label{maggen}
\psi_m(\bm{r}) = \{ \exp[-i \epsilon(1 + \cos K z) f^x] \}_{m, -1}
\Psi_{-1}(\bm{r}),
\end{equation}
where $\epsilon \ll 1$ is a constant.
The operator acting on $\Psi_{-1}$ in Eq.~(\ref{maggen}) generates magnons
propagating in the $\pm z$ directions, which is the long axis of the
oblate condensate.
Since the experimental results are extrapolated to $\epsilon \rightarrow
0$~\cite{Marti}, we typically take $\epsilon = 0.01$ in the numerical
calculations; this is small enough and the numerical results are
insensitive to the value of $\epsilon$.

We numerically solve Eq.~(\ref{GP}) using the pseudospectral
method~\cite{Recipes}.
The initial ground state $\Psi_{-1}$ is prepared by the imaginary-time
propagation method, in which $i$ on the left-hand side of Eq.~(\ref{GP})
is replaced by $-1$.
The convolution integral in the MDDI term is calculated using the fast
Fourier transform.
The numerical mesh is typically $256 \times 16 \times 512$.

\begin{figure}[tbp]
\includegraphics[width=8cm]{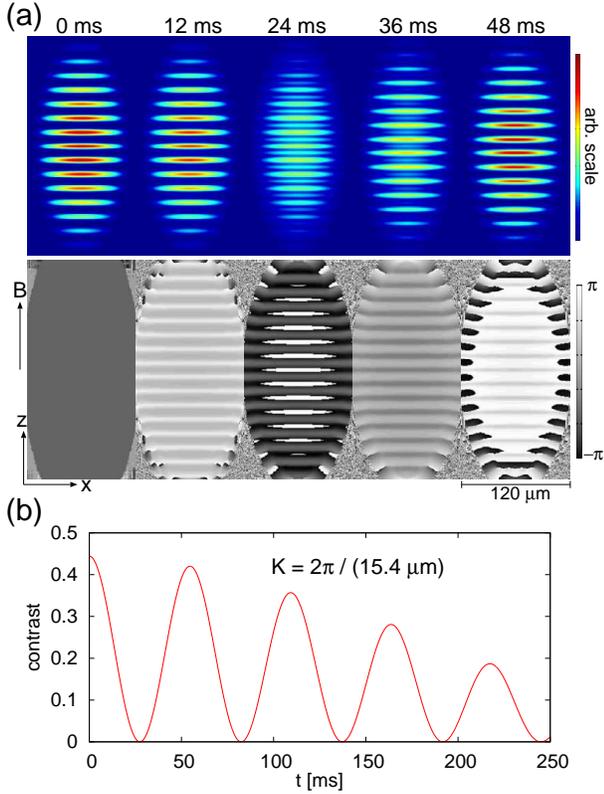}
\caption{
(color online) (a) Time evolution of the column density profile $\int
|\psi_0|^2 dy$ (upper panels) and cross-sectional phase profile ${\rm arg}
\psi_0(y = 0)$ (lower panels).
See Supplemental Material for a movie of the dynamics of the column
density profile~\cite{movie}.
(b) Time evolution of the fringe contrast.
The wave number of the excited magnons in (a) and (b) is $K = 2\pi / (15.4
\mu{\rm m}$).
}
\label{f:evolution}
\end{figure}
Figure~\ref{f:evolution}(a) shows the time evolution of the density and
phase profiles of the $m = 0$ component, which approximately represent the
magnon density and the tilting direction of the spin vector,
respectively.
Assuming a uniform system, the wave function of the $m = 0$ component
produced by Eq.~(\ref{maggen}) is $\psi_0 \propto 1 + \cos Kz$, which
evolves as $e^{-i \omega(0) t} + e^{-i \omega(K) t} \cos Kz$, giving the
density
\begin{equation} \label{psi0}
|\psi_0|^2 \propto 1 + \cos^2 Kz + 2 \cos \omega_0(K) t \cos Kz
\end{equation}
with $\omega_0(K) \equiv \omega(K) - \omega(0)$.
The fringes of $|\psi_0|^2$ in Fig.~\ref{f:evolution}(a) thus has a pitch
of $2\pi / K$ for $\cos \omega_0(K) t \simeq \pm 1$ ($t = 0$ and 48 ms)
and $2\pi / (2 K)$ for $\cos \omega_0(K) t \simeq 0$ ($t = 24$ ms).
Figure~\ref{f:evolution}(b) shows the fringe contrast defined by
$|\int e^{-i K z} D(z) dz|^2 / |\int D(z) dz|^2$, where $D(z) \equiv \int
|\psi_0|^2 dxdy$.
From Eq.~(\ref{psi0}), the fringe contrast becomes $(2/3)^2 \cos^2
\omega_0(K)t$ for a uniform system.
The decay in the oscillation amplitude in Fig.~\ref{f:evolution}(b) is
mainly due to the inhomogeneity of the system.

\begin{figure}[tbp]
\includegraphics[width=9cm]{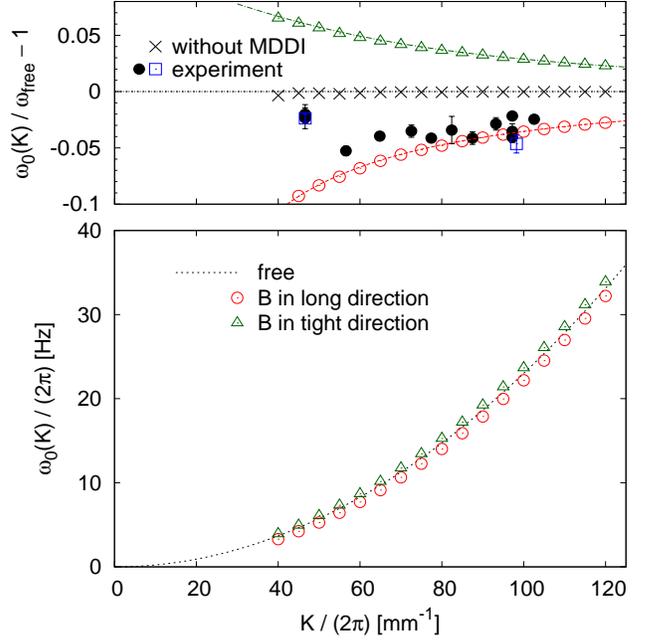}
\caption{
(color online) Dispersion relation of a magnon, where $\omega_0(K) \equiv
\omega(K) - \omega(0)$.
The red circles and green triangles are obtained by numerically solving
the GP equation, where in the former, the magnetic field is in the
direction of the long axis of the oblate condensate, and in the latter,
the magnetic field is in the direction of the tight axis.
The black crosses are the numerical result without the MDDI.
The black dotted curve shows the dispersion relation for a free atom.
The black solid circles and the blue squares are the experimental data,
where the former was obtained from a single tipping angle of the spin and
the latter was obtained from several tipping angles ~\cite{Marti}.
The red dashed and green dot-dashed curves are obtained from
Eq.~(\ref{u0dot}).
}
\label{f:dispersion}
\end{figure}
Fitting the fringe contrast to a function $[(2/3)^2 - at - bt^2] \cos^2
\omega t$ with fitting parameters $a$, $b$, and $\omega$~\cite{Note}, we
obtain the red circles in Fig.~\ref{f:dispersion}.
As observed in the experiment~\cite{Marti}, the dispersion curve lies
below that of a free atom (black dotted line in Fig.~\ref{f:dispersion}).
In the upper panel of Fig.~\ref{f:dispersion}, we find that the
experimental data are in reasonable agreement with our numerical result.
The crosses in Fig.~\ref{f:dispersion} are obtained by solving the GP
equation without the MDDI, and these almost coincide with the dispersion
relation of a free atom, indicating that there is little change in the
dispersion curve due to the finite-size effect of the trapped system.

To understand the frequency shift obtained above, we consider a linear
approximation of Eq.~(\ref{GP}).
Substituting $\psi_{-1} = \Psi_{-1} + \delta\psi_{-1}$, $\psi_0 =
\delta\psi_0$, and $\psi_1 = \delta\psi_1$ into Eq.~(\ref{GP}), and taking
the first order of $\delta\psi_m$, we obtain
\begin{eqnarray} \label{Bogo}
i \hbar \frac{\partial \delta\psi_0}{\partial t} & = & \left(
-\frac{\hbar^2}{2M} \nabla^2 + V \right) \delta\psi_0 
\nonumber \\
& & - g_d \int d\bm{r}'
\frac{1 - 3 \cos^2\theta}{|\bm{r} - \bm{r}'|^3} \Psi_{-1}^*(\bm{r}')
\Psi_{-1}(\bm{r}) \delta\psi_0(\bm{r}').
\nonumber \\
\end{eqnarray}
For a uniform system without the MDDI, Eq.~(\ref{Bogo}) reduces to $i
\hbar \partial \delta\psi_0 / \partial t = -\hbar^2 / (2M) \nabla^2
\delta\psi_0$, which gives the dispersion relation for a free atom,
$\omega(K) = \hbar K^2 / (2M)$.
To evaluate the frequency shift by the MDDI, we assume that $\delta\psi_0$
has the same profile as that of $\Psi_{-1}$:
\begin{equation} \label{delta0}
\delta\psi_0(\bm{r}, t) = e^{i\bm{K} \cdot \bm{r}} \Psi_{-1}(\bm{r})
u_0(t),
\end{equation}
where $\bm{K}$ is the wave vector of magnons.
Substituting Eq.~(\ref{delta0}) into Eq.~(\ref{Bogo}), multiplying by
$e^{-i\bm{K} \cdot \bm{r}} \Psi_{-1}^*$, and integrating with respect to
$\bm{r}$, we obtain
\begin{eqnarray} \label{u0dot}
i \hbar \dot{u}_0(t) & = & \biggl[ \frac{1}{N} \int d\bm{r} \Psi_{-1}
\left( -\frac{\hbar^2}{2M} \nabla^2 + V \right) \Psi_{-1}
+ \frac{\hbar^2 K^2}{2M} 
\nonumber \\
& & + \frac{4\pi g_d}{3 N} \int \frac{d\bm{k}}{(2\pi)^3} (1 - 3 \cos^2
\alpha) |\phi(\bm{k} - \bm{K})|^2 \biggr] u_0(t),
\nonumber \\
\end{eqnarray}
where $\alpha$ is the angle between $\bm{k}$ and the direction of the
magnetic field, and $\phi(\bm{k}) \equiv \int d\bm{r} \exp(-i \bm{k} \cdot
\bm{r}) |\Psi_{-1}|^2$.
The first term in the square bracket in Eq.~(\ref{u0dot}) is a constant
that does not contribute to $\omega_0(K)$.
The second term gives the dispersion relation of a free atom.
The $K$ dependence of the third term determines the MDDI effect on
$\omega_0(K)$.
The red dashed curve in Fig.~\ref{f:dispersion} shows the frequency shift
obtained from Eq.~(\ref{u0dot}), where we used the numerically obtained
$\Psi_{-1}$ to calculate the MDDI term in Eq.~(\ref{u0dot}).
The frequency shift obtained from Eq.~(\ref{u0dot}) is in good agreement
with that obtained from the full GP equation.

If we take a Gaussian wave function as $\Psi_{-1}$, we have
\begin{equation} \label{gaussian}
|\phi(\bm{k} - K \hat z)|^2 = N e^{-[d_x^2 k_x^2 + d_y^2 k_y^2 + d_z^2
(k_z - K)^2] / 2},
\end{equation}
where $\hat z$ is the unit vector in the $z$ direction, and $d_x$, $d_y$,
and $d_z$ are Gaussian widths.
Suppose that $d_z \gg K^{-1}$, Eq.~(\ref{gaussian}) is approximated by
$|\phi(\bm{k} - K \hat z)|^2 \propto \delta(k_z - K)$, and therefore, $1 -
3 \cos^2 \alpha \sim 1 - 3 K^2 / (K^2 + k_x^2 + k_y^2)$.
The MDDI integral in Eq.~(\ref{u0dot}) is thus a decreasing function of
$K$ and hence the frequency shift is negative~\cite{Note2}.

We propose to change the direction of the magnetic field from the long
axis to the tight axis of the oblate BEC (i.e., from the $z$ direction to
the $y$ direction).
The quantization axis of the spin is always taken to be in the direction
of the magnetic field; the spin is initially in the $-y$ direction.
The green triangles in Fig.~\ref{f:dispersion} are obtained by numerically
solving the GP equation with fitting of the fringe contrast, and the
green dot-dashed curve in Fig.~\ref{f:dispersion} is obtained from
Eq.~(\ref{u0dot}).
The dispersion relation deviates to the upper side of that of a free atom,
which indicates that the effective mass of a magnon decreases,
contrary to the case in which the magnetic field is in the direction of
the long axis.
From the upper panel of Fig.~\ref{f:dispersion}, we see that the effective
mass of a magnon is a few percent smaller than the mass of a free atom.
In a similar manner as the above Gaussian approximation, we have $1 - 3
\cos^2 \alpha \sim 1 - 3 k_y^2 / (K^2 + k_x^2 + k_y^2)$, and therefore the
MDDI integral in Eq.~(\ref{u0dot}) is an increasing function of $K$,
giving the positive frequency shift.

\begin{figure}[tbp]
\includegraphics[width=9cm]{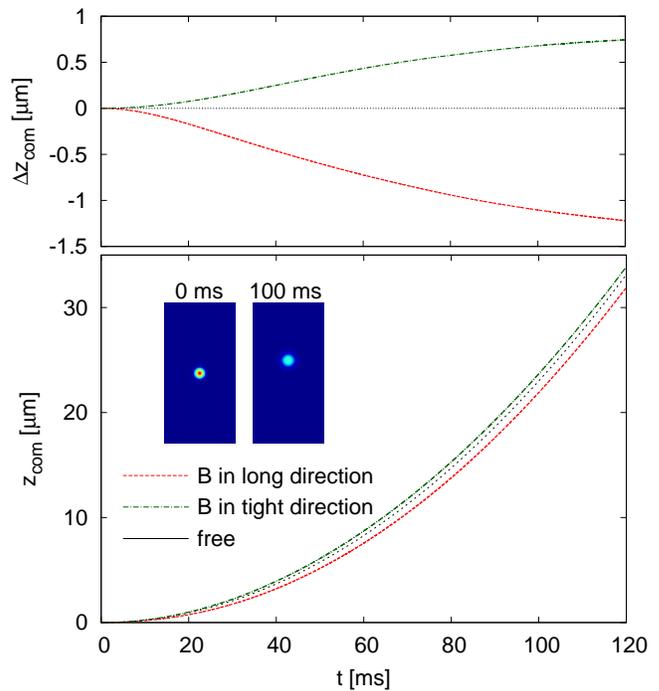}
\caption{
(color online) Time evolution of the position $z_{\rm com}$ of the center
of mass of locally excited magnons, where the Larmor frequency has a
gradient of 1 ${\rm kHz} / {\rm mm}$ in the $z$ direction.
The magnetic field is in the direction of the long axis (red dashed curve)
or in the direction of the tight axis (green dot-dashed curve).
The black dotted curve shows the motion of a free atom $z_{\rm free}$.
The upper panel shows $\Delta z_{\rm com} = z_{\rm com} - z_{\rm free}$.
The insets are snapshots of the column density profile of the $m = 0$
component at $t = 0$ and 100 ms, where the window is the same as in
Fig.~\ref{f:evolution}(a).
The dynamics of the column density profile is provided as Supplemental
Material~\cite{movie}.
}
\label{f:accel}
\end{figure}
In the experiment~\cite{Marti}, the effect of the heavy magnon mass on the
acceleration dynamics was not observed due to a systematic error.
To clarify the effect of the MDDI on the acceleration dynamics of magnons,
we investigate the same situation numerically.
The magnetic field gradient is applied along the long axis of the
condensate; this produces the gradient of the Larmor frequency,
$\partial_z f_{\rm Larmor}$.
In the initial state, localized magnons are excited as
\begin{equation} \label{maggen2}
\psi_m(\bm{r}) = \{ \exp[-i \epsilon e^{-(x^2 + z^2) / \sigma^2}] f^x
\}_{m, -1} \Psi_{-1}(\bm{r}),
\end{equation}
where $\epsilon = 0.01$ and $\sigma = 12$ $\mu{\rm m}$.
Figure~\ref{f:accel} shows the time evolution of the position of the
center of mass, $\bm{r}_{\rm com} = \int \bm{r} |\psi_0|^2 d\bm{r} / \int
|\psi_0|^2 d\bm{r}$; this exhibits the parabolic acceleration of magnons
in the direction of the field gradient, as observed in the experiment.
The deviations from the motion of a free atom are shown in the upper panel
of Fig.~\ref{f:accel}.
As expected, the acceleration is suppressed when the magnon mass
increases, and it is enhanced when the magnon mass decreases. 
We fit $z_{\rm com}(t)$ in Fig.~\ref{f:accel} to $h|\partial_z
f_{\rm Larmor}| t^2 / (2M^*)$ to estimate the effective mass $M^*$, where
we assume that the effective magnetic moment $\mu^*$ is the same as that
of a free atom $\mu$ and that the effective mass $M^*$ is averaged over
the density distribution and the momentum range.
The estimated effective mass is $M^* / M \simeq 1.05$ and $0.97$ for the
red dashed curve and the green dot-dashed curve in Fig.~\ref{f:accel},
respectively, which are consistent with the dispersion relation in
Fig.~\ref{f:dispersion}.

In conclusion, we have investigated the dynamics of magnons in a
polarized spin-1 $^{87}{\rm Rb}$ BEC using the mean-field theory with the
MDDI.
In the same situation as was used in the experiment~\cite{Marti}, we have
shown that the dispersion relation of a magnon lies below that of a free
atom, which agrees with the experiment.
We have confirmed that the deviation mainly arises from the MDDI.
We also proposed a simple modification to the experiment:
when the direction of the magnetic field is changed from the long axis to
the tight axis of the oblate condensate, the dispersion relation deviates
to the upper side, and the effective mass of a magnon decreases.
This effect can be observed with the current experimental precision.
We have shown that the change in the magnon mass affects the acceleration
dynamics of magnons in the field gradient.
The acceleration is suppressed (enhanced) when the magnon mass is
increased (decreased).
Since our method can both increase and decrease the magnon mass, the
difference in their acceleration dynamics may be detected even in the
presence of the systematic error of the experiment.

Since the MDDI shift is already detectable for $^{87}{\rm Rb}$, magnons in
BECs of $^{52}{\rm Cr}$, $^{164}{\rm Dy}$, and $^{168}{\rm Er}$ are
strongly affected by the MDDI.
For example, for $^{52}{\rm Cr}$, the ratio between the MDDI and kinetic
energies is $12^2 m_{\rm Rb} / m_{\rm Cr} \simeq 86$ times larger than the
present case.
The strong dipolar effect may exhibit a novel dynamics of magnons.

We thank G. Edward Marti for the experimental data and valuable comments.
This work was supported by JSPS KAKENHI Grant Number 26400414 and by
KAKENHI (No. 25103007, ``Fluctuation \& Structure'') from MEXT, Japan.


\begin{thebibliography}{99}
\bibitem{Note}
Since the initial one or two oscillations are used for fitting, $\omega$
obtained from the fitting is insensitive to the form of the fitting
function.
\end{thebibliography}
\end{document}